\documentclass[aps, pra, twocolumn, superscriptaddress]{revtex4-2}
\usepackage{graphicx}
\usepackage{amsmath}
\usepackage{amssymb}
\usepackage{bm}

\usepackage[colorlinks, linkcolor=blue, urlcolor=magenta, citecolor=blue]{hyperref}

\begin{document}

\title{Sensitivity Enhancement near High-Order Exceptional Points via Dissipative Couplings}
\author{Yuanjie Zhang}
\email[]{These authors contributed equally to this work}
\affiliation{Guangdong Provincial Key Laboratory of Quantum Metrology and Sensing \& School of Physics and Astronomy, Sun Yat-Sen University (Zhuhai Campus), Zhuhai 519082, China.}

\author{Jiaojiao Li}
\email[]{These authors contributed equally to this work}
\affiliation{Guangdong Provincial Key Laboratory of Quantum Metrology and Sensing \& School of Physics and Astronomy, Sun Yat-Sen University (Zhuhai Campus), Zhuhai 519082, China.}

\author{Zhihuang Luo}
\email{luozhih5@mail.sysu.edu.cn}
\affiliation{Guangdong Provincial Key Laboratory of Quantum Metrology and Sensing \& School of Physics and Astronomy, Sun Yat-Sen University (Zhuhai Campus), Zhuhai 519082, China.}


\begin{abstract}
High-order exceptional points (EPs) emerging in non-Hermitian systems have attracted broad interest for their significantly enhanced sensitivity to perturbations. However, quantum sensing schemes based on high-order EPs remain scarce, due to the experimental challenge of fine-tuning the system to such an extremely sensitive isolated point. Here we propose a four-channel dissipative coupling model that supports both fourth-order exceptional surfaces and second-order exceptional volumes. This non-Hermitian model can be realized in a thermal atomic system, and its complex energy spectra can be determined via electromagnetically induced transparency spectroscopy. The proposed model exhibits a characteristic fourth-order response to multiple physical quantities such as the laser detuning and the distance between optical channels, significantly surpassing the response of second-order EPs. We further reveal the sensitivity-robustness trade-off under experimental noise. Our work opens a route toward high-performance sensing leveraging higher-order EPs.
\end{abstract}

\maketitle


\section{Introduction} 

Non-Hermitian systems~\cite{Bender_2007,Ashida2020} have attracted considerable attention due to their distinctive properties, which markedly differ from those of Hermitian systems~\cite{Yokomizo2022}. Among their intriguing features, exceptional points (EPs)~\cite{Kato1966,Berry2004,Heiss2003,1998Diffraction,1999Phases,2017Non}, the non-Hermitian skin effect~\cite{Gui2024Localization,2020Topological,Zhang2021}, parity-time (PT) symmetry~\cite{1998Real,AHMED2001343,bender2002complex,Ozdemir2019Parity}, and non-trivial topology~\cite{2020Observation,2019Symmetry,
2019Anatomy} stand out as particularly remarkable. These unique characteristics have enabled novel applications including unidirectional invisibility~\cite{Lin2011Unidirectional,2025Nonreciprocity,2019Unidirectional}, single-mode lasing~\cite{Contractor2022,2018Topological}, and enhanced spontaneous emission~\cite{2016Enhanced,PhysRevLett.129.083602,
2021Control}. A particularly promising direction is to exploit the singular properties of EPs for designing ultralsensitive sensors~\cite{2014Enhancing,0Enhanced,Wiersig2016Sensors,2021High,RevModPhys.89.035002}. In contrast to the linear response observed in Hermitian systems~\cite{1984Diabolical}, operating near an $n$th-order EP enhances the
response of system to the $n$th root of the perturbation strength~\cite{2005Coupling,2017Exceptional}, thereby enabling exponential signal amplification.

Recently, EP-based sensors have been extensively explored across diverse physical platforms~\cite{Wu2025Recent,Wiersig2020Review}, demonstrating potential beyond conventional sensing paradigms. In optics, EPs have been successfully utilized in microcavity-based particle detection~\cite{Zhu2010Controlled,Wiersig2011Structure}, PT-symmetric microring resonators~\cite{0Enhanced,Peng2014Parity}, and topological waveguides~\cite{2023Higher}, achieving highly responsive detection of weak perturbations. In electronics, PT-symmetric wireless sensing systems~\cite{8466793,2018Generalized} and sensing circuit~\cite{Zhicheng0Enhanced} have further broadened their practical applicability. 
Nevertheless, the practical implementation of EP-based sensors faces several critical challenges. Intrinsic noise floors limit the ultimate detection sensitivity~\cite{2019Quantum,2018No}, while the system response is highly anisotropic to perturbation directions, often leading to mode destabilization~\cite{2021High,2022Highly}. Moreover, extreme sensitivity enhancement near high-order EPs makes it particularly difficult to prepare the system in an ideal initial state~\cite{2024Higher}. As a result, quantum sensing based on high-order EPs remains largely unexplored.

In this work, we propose a four-channel dissipative coupling model that supports rich exceptional structures across a broad range of controllable parameters, including second-order exceptional volumes (EVs) and fourth-order exceptional surfaces (ESs). These structures enable the system to be prepared at high-order EPs over a wide range of easily accessible control parameters to achieve high sensing performance, overcoming the limitation of conventional systems that rely on a single, isolated EP. 
We further demonstrate the feasibility of realizing this model in a thermal atomic system via electromagnetically induced transparency (EIT) and extracting its complex energy spectra by sweeping an applied magnetic field during probe transmission measurements ~\cite{2016Anti,2023Topological}.
Although the fourth-order EP offers higher sensitivity enhancement, it is also more susceptible to noise, in contrast to the second-order EP.
The proposed scheme based on high-order EPs can provide exponentially enhanced response to multiple physical quantities such as the laser detuning and the distance between optical channels, opening a new avenue for developing ultrasensitive sensors.

\section{\label{sec:level2}Theoretical Framework}

\begin{figure}
   \centering
   \includegraphics[width=0.95\linewidth]{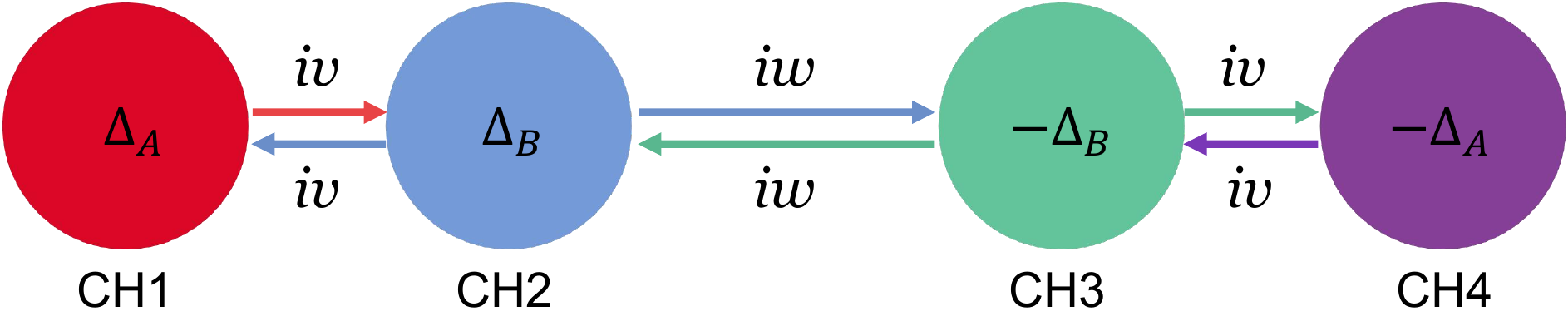} 
   \caption{Schematic of a four-channel dissipative coupling model. Four optical channels with eigenfrequencies ($\Delta_A$,$\Delta_B$,$-\Delta_B$,$-\Delta_A$) are arranged in a chain and coupled dissipatively to their nearest neighbors with coupling strengths $(iv, iw, iv)$.}
   \label{fig: model}
\end{figure}

We begin by considering a four-channel dissipative coupling model as shown in Fig. \ref{fig: model}, whose dynamics are governed by the following non-Hermitian Hamiltonian
\begin{eqnarray}\label{eq: Hamiltonian}
    H=\left(\begin{array}{cccc}
 \Delta_A&iv&0&0\\iv&\Delta_B&iw&0\\0&iw&-\Delta_B&iv\\0&0&iv&-\Delta_A
    \end{array}\right),
\end{eqnarray}
where $[\Delta_A$,$\Delta_B$,$-\Delta_B$,$-\Delta_A]$ represent the eigenfrequencies, and $iv$ and $iw$ (here $v,w$ are real) denote the coupling strengths. The Hamiltonian \eqref{eq: Hamiltonian} can be realized in a thermal atomic system (see Sec. \ref{sec: proposal} for the details). 

The Hamiltonian in Eq.~\eqref{eq: Hamiltonian} is shown to exhibit anti-PT symmetry, due to $\{PT, H\} = 0 $. Its eigenvalues $\lambda_j$s for $j= 1, 2, 3, 4$ are given by
\begin{equation}\label{eq: eigenvalue}
    \lambda_j=\pm \sqrt{(\Phi\pm \sqrt{\Pi})/2}, 
\end{equation}
where $\Phi = \Delta_A^2 + \Delta_B^2 - 2v^2 - w^2$, and $\Pi = \Phi^2 - 4\Xi_1\Xi_2$
with $\Xi_1 = \Delta_A\Delta_B + v^2 + w\Delta_A$ and $\Xi_2 = \Delta_A\Delta_B + v^2 - w\Delta_A$. The corresponding eigenvectors can be obtained as 
\begin{equation}\label{eq: eigenvector}
    \bm{V}_{\lambda_j} = \left [i\frac{M_j}{w(\Delta_A - \lambda_j)}, -\frac{M_j}{wv}, -i\frac{\Delta_A + \lambda_j}{v}, 1\right ]^T/N_j.
\end{equation}
Here $M_j = (\Delta_A + \lambda_j)(\Delta_B + \lambda_j) + v^2$ and $N_j$ denotes the normalization factor for each $\lambda_j$.

\begin{figure}
   \centering
   \includegraphics[width=0.5\textwidth]{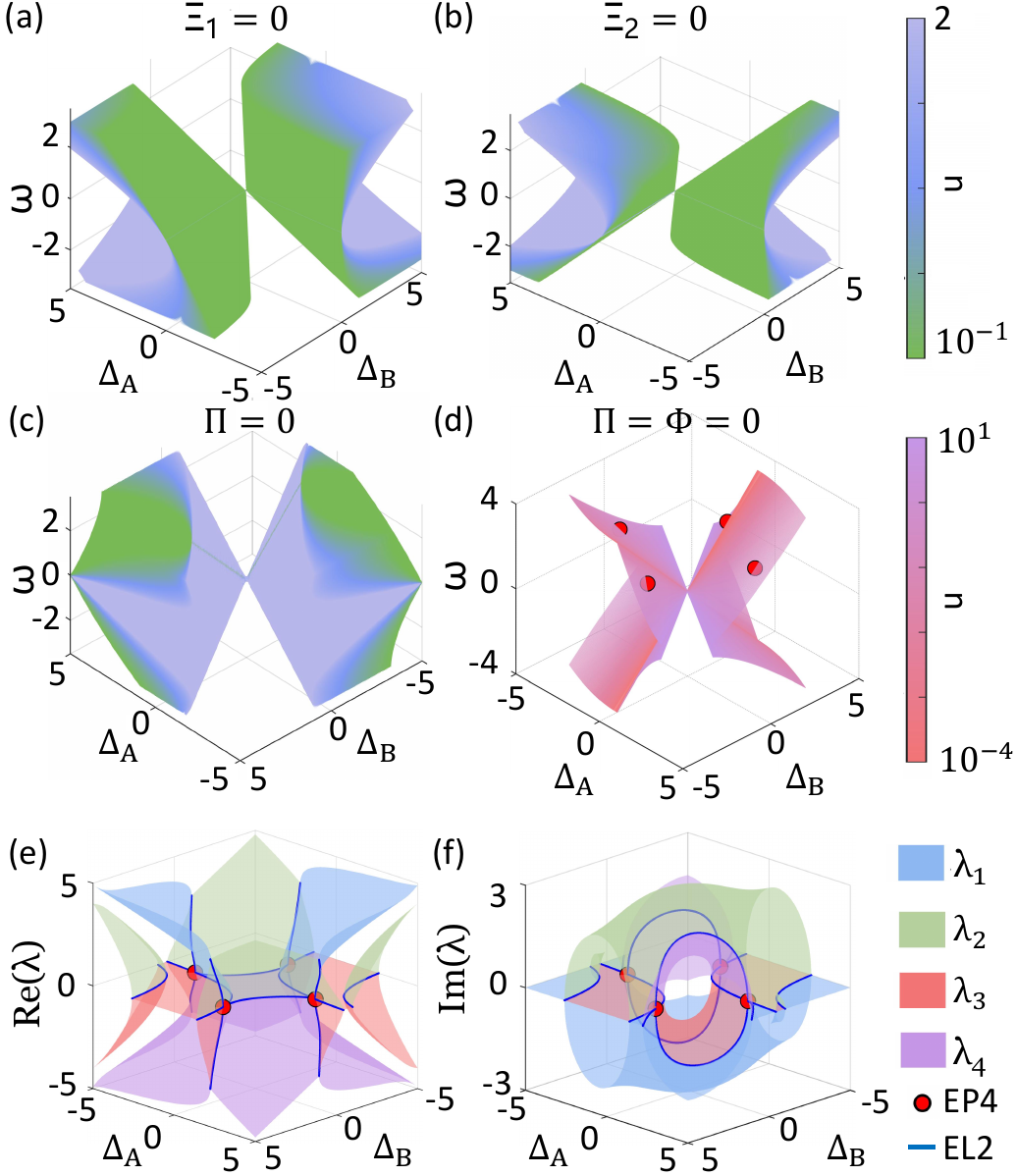};
   \caption{(a-c) Second-order EVs consisting of three hypersurfaces: $\Xi_1 =0$, $\Xi_2 = 0$ and $\Pi = 0$ defined in the 4D parameter space $\{\Delta_A, \Delta_B, v, w\}$, where $v = u w$. The colorbar represents the value of $u$.
(d) Fourth-order ESs determined by the intersection of two hypersurfaces $\Pi = 0$ and $\Phi = 0$. The red points mark the locations of fourth-order EPs, corresponding to the fixed parameters used in (e) and (f). 
(e) and (f) Real and imaginary parts of complex eigenvalues $\lambda_j$ ($j=1,2,3,4$) for fixed $v = \sqrt{3}$ and $w = 2$. Red points and blue curves indicate fourth-order EPs (EP4s) and second-order exceptional lines (EL2s), respectively. 
}
   \label{fig: EP}
\end{figure}

To analyze the geometric structure of high-order EPs in a 4D parameter space $\{\Delta_A, \Delta_B, v, w\}$, we first use the resultant method~\cite{article} to study algebraic multiplicities of the characteristic polynomial, i.e., \( f(\lambda) = |\lambda I - H| = \lambda^4 - \Phi \lambda^2 + \Xi_1\Xi_2 = 0 \). Constructing the Sylvester matrices \( \text{Syl}(f, f^{(i)}) \) from the coefficients of $f(\lambda)$ and computing their determinants yield the conditions for algebraic multiplicities: $\det\left [\text{Syl}(f,f^{(1)})\right ]=16\Pi^2\Xi_1\Xi_2 = 0, \det\left [\text{Syl}(f,f^{(2)})\right ]=16(5\Phi^2 - 36\Xi_1\Xi_2)^2 = 0, \det\left [\text{Syl}(f,f^{(3)})\right ]=331776\Xi_1\Xi_2 = 0$, where $f^{(i)}$ stands for the $i$th derivative. From these conditions, it can be readily found that second-order EPs emerge in three hypersurfaces defined by $\Xi_1 =0$, $\Xi_2 = 0$ and $\Pi = 0$, as shown in Figs. \ref{fig: EP}(a) - \ref{fig: EP}(c). At points in these hypersurfaces, any two of the four eigenvalues coincide, causing their corresponding eigenvectors to coalesce, according to Eqs. \eqref{eq: eigenvalue} and \eqref{eq: eigenvector}. In order to visualize the second-order EVs in the 4D parameter space, we set \( v = u w \) and vary the value of $u$, as plotted in Figs. \ref{fig: EP}(a) - \ref{fig: EP}(c). 

Furthermore, when $\Pi = \Phi = 0$, all resultants of the Sylvester matrices $\text{Syl}(f, f^{(i)})$ for $i = 1, 2, 3$ vanish simultaneously. The intersection of two hypersurfaces $\Pi = 0$ and $\Phi = 0$ forms a 2D manifold, as shown in Fig. \ref{fig: EP}(d), where the characteristic polynomial $f(\lambda)$ has algebraic multiplicity four. At points on this manifold, all four eigenvalues $\lambda_j$s become degenerate, and their eigenvectors $\bm{V}_{\lambda_j}$ coalesce into a single one. Consequently, the manifold in Fig. \ref{fig: EP}(d) corresponds to a fourth-order ES. Note that there is no third-order EPs because the condition of $\det\left [\text{Syl}(f,f^{(3)})\right ] = 0$ is automatically satisfied if $\det\left [\text{Syl}(f,f^{(i)})\right ] = 0$ for $i=1,2$.

Figures \ref{fig: EP}(e) and \ref{fig: EP}(f) show the real and imaginary parts of complex eigenvalues $\lambda_j$ for $j=1,2,3,4$, where $v = \sqrt{3}$ and $w = 2$ are fixed. In the 2D parameter space of $\{\Delta_A, \Delta_B\}$, the four energy bands exhibit pairwise mirror symmetry about the $\lambda=0$ plane, resulting in symmetric EPs. Blue curves indicate second-order exceptional lines (EL2s), which arise from the degeneracies of any two eigenvalues, as illustrated in Figs. \ref{fig: EP}(e) and \ref{fig: EP}(f). The intersections of EL2s further form the fourth-order EPs (EP4s), in agreement with the red points marked in Figs. \ref{fig: EP}(d).
 
\section{Sensitivity at high-order EPs}

\begin{figure*}
   \centering
   \includegraphics[width=\linewidth]{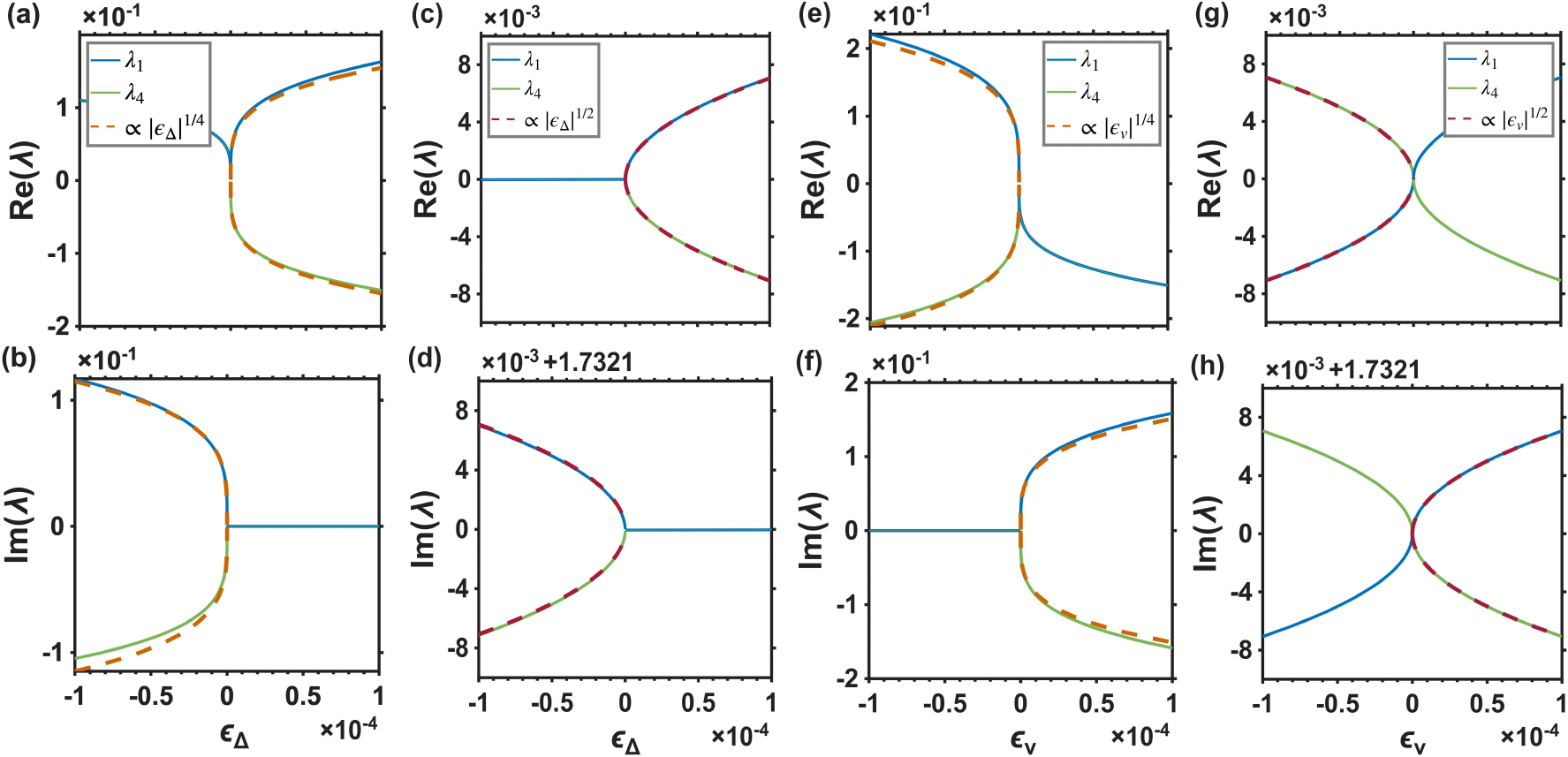} 
   \caption{Real and imaginary parts of the maximum and minimum eigenvalues $\lambda_1$ and $\lambda_4$ versus the eigenfrequency disturbance $\epsilon_{\Delta}$ (a-d) and the coupling disturbance $\epsilon_v$ (e-h) near EPs. In (a), (b), (e), and (f), the system is tuned to a fourth-order EP with parameters $\Delta_A = 3$, $\Delta_B = 1$, $v = \sqrt{3}$, $w = 2$. In (c), (d), (g), and (h), the system operates at a second-order EP with parameters $\Delta_A = 0$, $\Delta_B = 2$, $v = \sqrt{3}$, $w = 2$. The red and dotted curves represent the power-law fittings of the eigenvalue splittings near the fourth-order EP ($\propto |\epsilon_{\Delta,v}|^{1/4}$) and the second-order EP ($\propto |\epsilon_{\Delta, v}|^{1/2}$).}
   \label{fig: eigenvalue}
\end{figure*}

A slight deviation of any parameter drives the system from the EP into either the anti-PT-symmetric or symmetry-broken phase, accompanied by significant changes in eigenvalues. 
If we apply a disturbance \( \epsilon \) to the sensor model near an $n$th-order EP, the resulting energy splitting follows \( \Delta\lambda \propto \sqrt[n]{\epsilon} \)~\cite{0Enhanced}. 
Although a higher-order EP promises in principle greater sensitivity enhancement, it demands increasingly stringent experimental control to prepare the system precisely at such an EP.
Rather than requiring fine-tuning to an extremely sensitive isolated EP, the coexistence of second-order EVs and fourth-order ESs in our non-Hermitian model enables the extreme sensitivity enhancement to be achieved more readily over a broad range of control parameters in experiments.

As examples, we consider two types of disturbances $\epsilon_{\Delta}$ and $\epsilon_v$, which are applied to the eigenfrequency $\Delta_A$ on CH1 and the coupling $v$ between CH1 and CH2, as shown in Fig. \ref{fig: model}, respectively.
In the presence of disturbances, the eigenvalue splittings follow a power-law scaling with respect to the disturbance strength, characterized by exponents $1/4$ for the fourth-order EP (see Figs. \ref{fig: eigenvalue}(a), \ref{fig: eigenvalue}(b), \ref{fig: eigenvalue}(e) and \ref{fig: eigenvalue}(f)) and $1/2$ for the second-order EP (see Figs. \ref{fig: eigenvalue}(c), \ref{fig: eigenvalue}(d), \ref{fig: eigenvalue}(g) and \ref{fig: eigenvalue}(h)). 
The eigenvalue splittings near the fourth-order EP are around one order of magnitude greater than those near the second-order EP. The numerical results match well with theoretical predictions.

\begin{figure}
   \centering
   \includegraphics[width=0.5\textwidth]{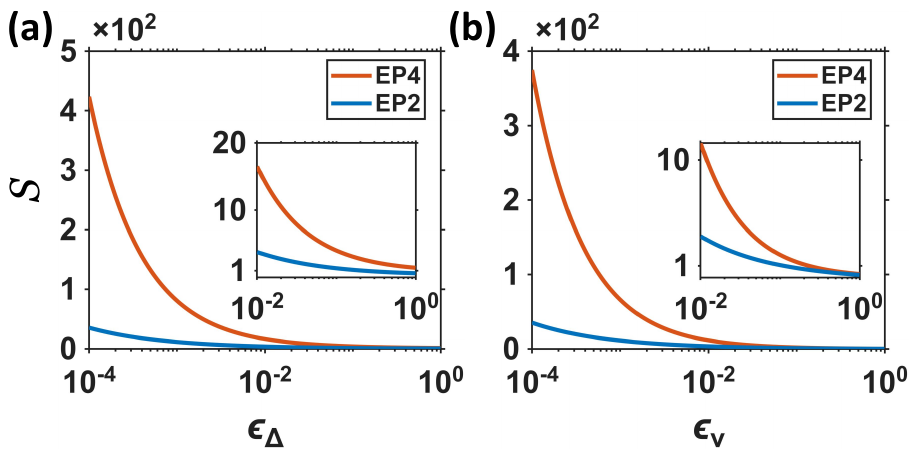}   
   \caption{Sensitivity enhancement factor $S$ as a function of the applied eigenfrequency disturbance $\epsilon_{\Delta}$ (a) and the coupling disturbance $\epsilon_v$ (b). The red and blue solid curves represent the sensitivity enhancement factors near the fourth- and second-order EPs, respectively.}
   \label{fig: sensitivity}
\end{figure}

To evaluate the sensing performance of our non-Hermitian model, we define the enhancement factor as
the ratio of the change in the real part of the largest eigenvalue to the applied disturbance strength $\epsilon_{\Delta}$ or $\epsilon_v$,
\begin{equation}
    S=\left |\frac{\partial \text{max}(\text{Re}\lambda_i)}{\partial\epsilon_{\Delta ,v}}\right |.
\end{equation}
The results are plotted in Figs. \ref{fig: sensitivity}(a) and \ref{fig: sensitivity}(b), showing that in the presence of EPs, the sensitivity enhancement factors increase for small values of $\epsilon_{\Delta}$ and $\epsilon_v$. In particular, 
the sensitivity enhancement near a fourth-order EP (up to hundreds of times) exceeds that of a second-order EP by more than an order of magnitude, and those of linear sensors ($\Delta\lambda\propto \epsilon$) by more than three orders of magnitude, when $\epsilon_{\Delta}$ and $\epsilon_v$ are below $10^{-4}$. While the values of $S$ for the second-order EPs become comparable to those of linear sensors, the sensitivity enhancements near the fourth-order EPs remain up to ten times for disturbances around $10^{-2}$ , as shown in the insets of Figs. \ref{fig: sensitivity}(a) and \ref{fig: sensitivity}(b).
The advantage of higher-order EP sensors gradually diminishes as the disturbance strength further increases. However, precision measurement is mainly concerned with the detection of exceedingly weak physical signals. In this regard, the fourth-order EP–based sensing scheme we propose provides a valuable approach for achieving enhanced sensitivity.
Additionally, the eigenfrequency $\Delta_A$ and the coupling $v$, related to the physical quantities like the detuning of optical channels and the distance between optical channels in our proposed experimental scheme (see Sec. \ref{sec: proposal}), can be measured with extremely high sensitivity.

\section{\label{sec: proposal} Towards experimental realization}

\begin{figure}
   \centering
   \includegraphics[width=0.48\textwidth]{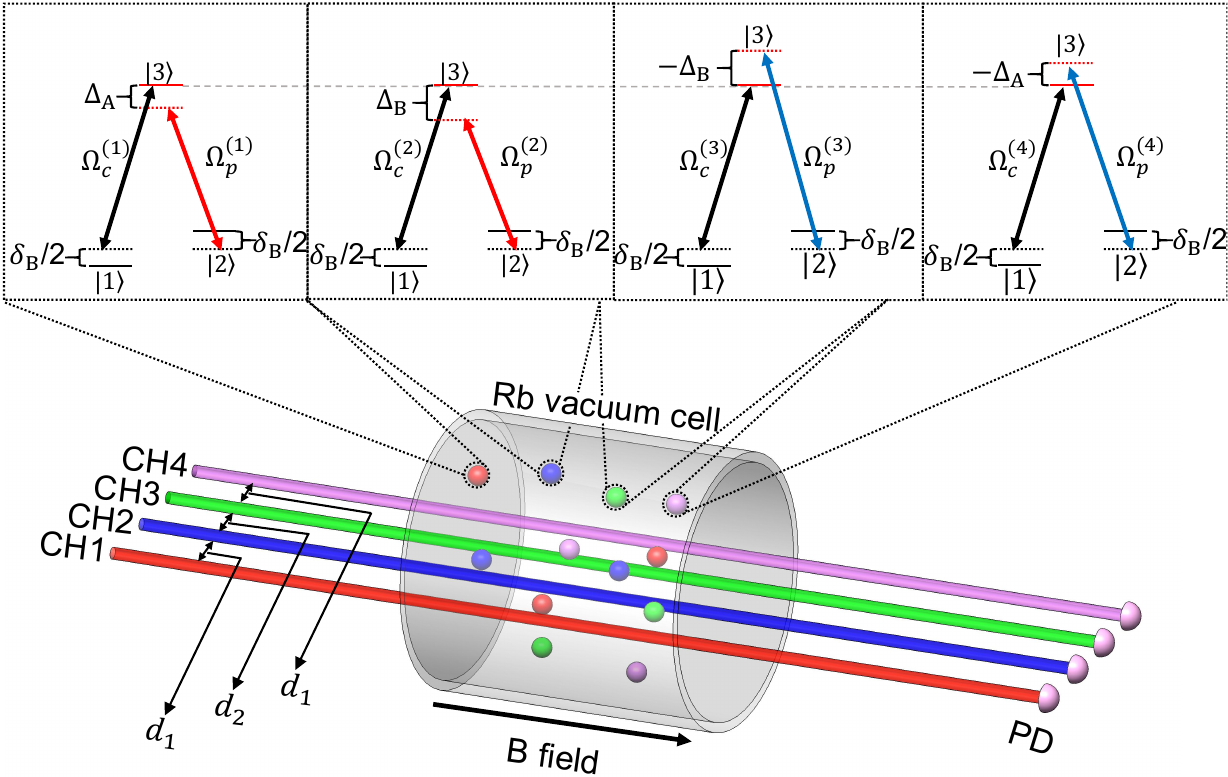} 
   \caption{Scheme for implementing a fourth-order non-Hermitian model using a thermal $^{87}$Rb atomic-vapor cell. Four optical channels are arranged in a chain with alternating spacings $(d_{1},d_{2},d_{1})$, each consisting of a spatially overlapped strong control beam and a weak probe beam. Within each channel, the atoms realize a standard $\Lambda$-type EIT involving two ground states $|1\rangle$ and $|2\rangle$ and an excited state $ |3\rangle $. The strong control fields with Rabi frequencies $\Omega_c^{(i)}$ for $i=1, 2, 3, 4$ drive the transitions $|1\rangle \to |3\rangle$, while the weak probe fields with Rabi frequencies $\Omega_p^{(i)}$ are on near resonance with the transitions $|2\rangle \to |3\rangle$. The frequency differences between the control and probe fields are $\Delta_A, \Delta_B, -\Delta_B, -\Delta_A$ for channels $1$ to $4$, respectively. The ground-state coherences generated by the EITs in  different optical channels are effectively coupled via the thermal motion of atoms. By sweeping a homogeneous magnetic field $B$, the output probe transmission spectra can be measured.}
   \label{fig: setup}
\end{figure}

We now construct a multichannel dissipative coupling model based on EIT~\cite{2016Anti,2023Topological}, as illustrated in Fig.~\ref{fig: setup}. Four optical channels are arranged in a chain with alternating spacings $(d_1, d_2, d_1)$. Each channel consists of a strong control field and a weak probe field, both collinearly propagating through a thermal $^{87}$Rb vacuum vapor cell. We employ a standard $\Lambda$-type EIT configuration in a three-level atomic system. The two ground states are Zeeman sublevels $|1\rangle = |F=2, m_F = 0\rangle$ and $|2\rangle = |F=2, m_F = 2\rangle$, and the excited state is $|3\rangle = |F=1, m_F =1\rangle$ of the $^{87}$Rb D1 line. The strong control field with Rabi frequency $\Omega_c$ is resonant with the $|1\rangle \to |3\rangle$ transition, while the weak probe field with Rabi frequency $\Omega_p$ is near-resonant with the transition $|2\rangle \to |3\rangle$. The frequency differences between the control and probe fields are set to \( (\Delta_A, \Delta_B, -\Delta_B, -\Delta_A) \) for channels $1$ to $4$, respectively, as indicated in Fig. \ref{fig: setup}.

Since the lifetime of excited-state coherence is much shorter than that of ground-state coherence, and $\Omega_c^{(i)}\gg \Omega_p^{(i)}$, we adiabatically eliminate the excited state and describe the system evolution using ground-state coherences $\rho_{12}^{(i)}$ for $i = 1, 2, 3, 4$. 
Owing to the thermal motion of atoms, the ground-state coherences associated with four optical channels can be effectively coupled. Its dynamics of ground-state coherences is expressed as (see Appendix \ref{sec: derivationofnonhermitianhamiltonian} for the details)
\begin{equation}\label{eq: groundstatecoherence}
\bm{\dot{\rho}_{12}} = i\left [(\delta_B + i\gamma_{12}^{\text{eff}})I - H_{\text{eff}}\right ] \bm{\rho_{12}} - \bm{P},
\end{equation} 
where the vector of the ground-state coherences is defined as $\bm{\rho_{12}} = [\rho_{12}^{(1)}, \rho_{12}^{(2)}, \rho_{12}^{(3)}, \rho_{12}^{(4)}]^T$, the $i$-th component of $\bm{P}$ is $ P_i = \frac{\Omega_c^{(i)*} \Omega_p^{(i)}}{\gamma_{23}} $, $\gamma_{12}^{\text{eff}}$ denotes the effective decoherence rate of the ground state, and $\gamma_{23}$ represents the decay rate of the coherence between states $|2\rangle$ and $|3\rangle$. This effective Hamiltonian $H_{\text{eff}}$ in Eq. \eqref{eq: groundstatecoherence} matches our non-Hermitian model \eqref{eq: Hamiltonian}, by choosing a fixed time $t$ and tuning the applied magnetic field to zero during the EIT spectrum measurements (see the derivation in Appendix \ref{sec: derivationofnonhermitianhamiltonian}), and thus exhibits rich structures featuring fourth-order ESs and second-order EVs.

Experimentally, the EIT spectrum can be obtained by measuring the transmitted weak-probe field after the vapor cell. The probe transmission for the $i$-th channel is $T_p^{(i)}=\exp[-\alpha L \operatorname{Im}(\rho_{23}^{(i)})/\Omega_p^{(i)}]$, which is determined by the optical coherence $\rho_{32}^{(i)}$. Here $ L $ is the length of the vacuum cell, and $ \alpha = 2n\mu_{0}^{2}/\lambda\epsilon_{0}\hbar $, with the atomic number density $n$, dipole moment $ \mu_0 $, optical wavelength $ \lambda $, and vacuum permittivity $ \epsilon_0 $.
Under the steady-state approximation, the optical coherence takes the form $\rho_{23}^{(i)}=i(\Omega_p^{(i)}+\Omega_c^{(i)}\rho_{12}^{(i)})/\gamma_{32}$ (see Appendix \ref{sec: derivationofnonhermitianhamiltonian} for the details), while the ground-state coherence is given by
\begin{equation}\label{eq: steadyrho12}
    \bm{\rho}_{12} = -iM^{-1} \bm{P},
\end{equation}
with $ M = (\delta_B + i\gamma_{12}^{\text{eff}})I - H_{\text{eff}} $, according to Eq. \eqref{eq: groundstatecoherence}. 

The spectrum of $H_{\text{eff}}$ can be detected by sweeping the $\delta_B$ 
during probe transmission measurements.
Let $\lambda_j^{\prime}$ and $|\psi_j\rangle$ be the eigenvalues and their corresponding eigenvectors of $H_{\text{eff}}$, satisfying $H_{\text{eff}}|\psi_j\rangle = \lambda_j^{\prime}|\psi_j\rangle$ for $j=1,2,3,4$. 
The input $\bm{P}$ can be prepared into an eigenvector $|\psi_j\rangle$ of $H_{\text{eff}}$ by adjusting $\Omega_p^{(i)}$ under the condition of $\Omega_p^{(i)}\ll \Omega_c^{(i)}$. Substituting this into Eq. \eqref{eq: steadyrho12} yields $\bm{\rho}_{12}=-i(\delta_B + i\gamma_{12}^{\text{eff}} - \lambda_j^{\prime})^{-1}|\psi_j\rangle$.
To probe the eigenvalue $\lambda_j^{\prime}$, we compare with the case where $H_{\text{eff}}$ is absent, i.e., only the probe fields with the same single-detuning $\delta_B/2$ is on, while all control fields are kept off.
In that case, Eq. \eqref{eq: steadyrho12} reduces to $\bm{\rho}_{12}^{0}=-i\left(\delta_B + i\gamma_{12}^{\text{eff}}\right)^{-1}|\psi_j\rangle $. 
Thus, for all channels $j = 1,2,3,4$, the eigenvalue can be obtained via
\begin{equation}
    \lambda_j^{\prime} = (\delta_B+i\gamma_{12}^{\text{eff}})(\rho_{12}^{(j)}-\rho_{12}^{0,(j)})/\rho_{12}^{(j)}.
\end{equation}
When the applied magnetic field is swept to zero ($\delta_B=0$), the effective Hamiltonian coincides exactly with the non-Hermitian Hamiltonian in Eq. \eqref{eq: Hamiltonian}. Consequently, one can extract the eigenvalues $\lambda_j$ ($j = 1, 2, 3, 4$) of the target system.


Note that in the experiment, the couplings beyond the nearest-neighbor (NN) channels such as the next-nearest-neighbor (NNN) and the third-nearest-neighbor (TNN) couplings cannot be ignored, which can be largely suppressed by introducing a ring-shaped pumping beam close to the cell wall (see Appendix \ref{sec: MCS}). We analyze the robustness of a fourth-order and a second-order EPs against the noise arising from the NNN and TNN couplings (see Appendix \ref{sec: noise}).

\section{Conclusions and discussion}

In summary, we have proposed a non-Hermitian model that exhibits rich exceptional structures, including second-order EVs and fourth-order ESs. Rather than requiring fine-tuning to a fixed EP, the coexistence of second-order EVs and fourth-order ESs in our model enables the extreme sensitivity enhancement to be achieved more readily across a broad range of experimentally accessible control parameters. 
In particular, near a fourth-order EP the sensitivity enhancement can reach up to hundreds of times, orders of magnitude larger than that near a second-order EP, owing to the power-law scaling $\Delta\lambda \propto \epsilon^{1/n}$. 
We have further shown that the proposed non-Hermitian Hamiltonian can be realized using a thermal $^{87}$Rb atomic-vapor cell with four dissipatively coupled optical channels based on EIT. In this platform, the spectrum of desired Hamiltonian can be extracted from the EIT spectra by sweeping the applied magnetic field and measuring the probe transmission.
Finally, the generality of our framework suggests that it can be extended to a wide range of physical platforms such as optical microcavities, superconducting qubits, and electric circuits, thereby enabling enhanced sensitivity measurements of diverse physical quantities.

\section*{Acknowledgements}
This work was supported by the National Natural Science Foundation (Grant No. 11805008), Guangdong Basic and Applied Basic Research Foundation (Grant No. 2024A1515011406), Fundamental Research Funds for the Central Universities, Sun-Yat-Sen University (Grant No. 23qnpy63), Guangdong Provincial Key Laboratory (Grant No. 2019B121203005).

 \appendix

\section{\label{sec: derivationofnonhermitianhamiltonian}Derivation of the effective non-Hermitian Hamiltonian}

As discussed in the main text, we construct four spatially separated optical channels arranged in a chain passing through a $^{87}$Rb atomic-vapor cell. In each optical channel, the atoms interact with a strong control field and a weak probe field. The control field with Rabi frequency $ \Omega_c $ drives the transition $ |1\rangle \to |3\rangle $, while the probe field with Rabi frequency $ \Omega_p $ drives the transition $ |2\rangle \to |3\rangle $, thereby forming a standard $ \Lambda $-type EIT configuration. The free Hamiltonian with three levels can be written as
\begin{equation}
H_0 = \hbar\omega_{1}|1\rangle\langle 1| + \hbar\omega_{2}|2\rangle\langle 2| + \hbar\omega_3|3\rangle\langle 3|.
\end{equation}
Here, $ |1\rangle $ and $ |2\rangle $ correspond to the Zeeman sublevels of the ground states of the $^{87}$Rb atom, specifically $|F=2, m_F = 0\rangle$ and $|F=2, m_F=2\rangle$, $ |3\rangle $ denotes the Zeeman sublevel of the excited state, $|F=1,m_F = 1\rangle$ in the $^{87}$Rb D1 line, and $\omega_j$s for $j= 1,2,3$ represent the eigenfrequencies of these energy levels. 

Under the optical rotating-wave approximation, the Hamiltonian in the presence of light-atom interactions takes the form ($ \hbar = 1 $):
\begin{equation}
    H_{\text{RWA}} = 
    \begin{pmatrix}    
\omega_{1} & 0 & -\dfrac{\Omega_c}{2}e^{-i\omega_c t} \\[6pt]
0 & \omega_{2} & -\dfrac{\Omega_p}{2}e^{-i\omega_p t} \\[6pt]
-\dfrac{\Omega_c^*}{2}e^{i\omega_c t} & -\dfrac{\Omega_p^*}{2}e^{i\omega_p t} & \omega_3
\end{pmatrix},
\end{equation}
where $ \omega_p $ and $ \omega_c $ denote the frequencies of the probe and control fields, respectively.
Furthermore, by performing an unitary transformation
\begin{equation}
    U_R = 
\begin{pmatrix}
e^{i\omega_c t} & 0 & 0 \\
0 & e^{i\omega_p t} & 0 \\
0 & 0 & 1
\end{pmatrix},
\end{equation}
and ignoring the unit matrix term $\omega_3 I $ that do not affect the dynamical evolution,
we obtain the time-independent Hamiltonian in the rotating frame,
\begin{equation}
\begin{aligned}
H_R &= U_R H_{\text{RWA}} U_R^\dagger - i U_R \frac{\partial U_R^\dagger}{\partial t} - \omega_3 I \\
&=  \begin{pmatrix}
-\Delta_c & 0 & -\dfrac{\Omega_c}{2} \\[6pt]
0 & -\Delta_p & -\dfrac{\Omega_p}{2} \\[6pt]
-\dfrac{\Omega_c^*}{2} & -\dfrac{\Omega_p^*}{2} & 0
\end{pmatrix}.
\end{aligned}
\end{equation}
Here, $ \Delta_c = \omega_c - \omega_{13}$ and $ \Delta_p = \omega_p - \omega_{23}$ are the single-photon detunings of the control and probe fields, respectively, and $\omega_{ij} = \omega_i - \omega_j$ stands for the frequency of the corresponding transition between the states $|i\rangle$ and $|j\rangle$.

In the presence of decoherence, the dynamical evolution of atomic density matrix $\rho^{(i)}$ for the $i$-th independent channel is described by the optical Bloch equation \cite{Scully_Zubairy_1997}
\begin{equation}\label{eq: MB}
    \dot{\rho}^{(i)}=-i[H_R^{(i)},\rho^{(i)}]+\mathcal{L}_{\Gamma}[\rho^{(i)}]
\end{equation}
where the superoperator $\mathcal{L}_\Gamma$ includes all decoherence effects. According to Eq. \eqref{eq: MB}, the equations of the atomic density matrix elements $ \rho_{ij} $ take the following form:
\begin{equation}
\left \{
    \begin{aligned}
  \dot{\rho}_{12}^{(i)} =& -\left (\gamma_{12} + i\delta^{(i)}\right )\rho_{12}^{(i)} + i\Omega_c^{(i)^*}\rho_{32}^{(i)} - i\Omega_p^{(i)}\rho_{13}^{(i)} \\
 \dot{\rho}_{31}^{(i)} =& -\left (\gamma_{13} - i\Delta_c^{(i)}\right )\rho_{31}^{(i)} + i\Omega_p^{(i)}\rho_{21}^{(i)} - i\Omega_c^{(i)}\left (\rho_{33}^{(i)} - \rho_{11}^{(i)}\right ) \\
\dot{\rho}_{32}^{(i)} =& -\left (\gamma_{23} - i\Delta_p^{(i)}\right )\rho_{32}^{(i)} + i\Omega_c^{(i)}\rho_{12}^{(i)} - i\Omega_p^{(i)}\left (\rho_{33}^{(i)} - \rho_{22}^{(i)}\right )\\
\dot{\rho}_{11}^{(i)}=&-i\Omega_{c}^{(i)}\rho_{13}^{(i)}+i\Omega_{c}^{(i)^{*}}\rho_{31}^{(i)}+\gamma\rho_{33}^{(i)}-\gamma^{\prime}\left(\rho_{11}^{(i)}-\rho_{22}^{(i)}\right) \\           
    \dot{\rho}_{22}^{(i)} = &-i\Omega_{p}^{(i)}\rho_{23}^{(i)}+i\Omega_{p}^{(i)^{*}}\rho_{32}^{(i)}+\gamma\rho_{33}^{(i)}+\gamma^{\prime}\left(\rho_{11}^{(i)}-\rho_{22}^{(i)}\right) \\
    \rho_{33}^{(i)}=&1 -\rho_{11}^{(i)} - \rho_{22}^{(i)}.
\end{aligned}
\right.
\end{equation}
Here, $\gamma_{ij}$ represents the decay rate of the coherence between states $ |i\rangle $ and $ |j\rangle $, $\gamma$ is the decay rate of the excited state, $\gamma^{\prime}$ stands for the decay rate of the ground-state population difference, and $\delta^{(i)} = \Delta_c^{(i)} - \Delta_p^{(i)}$ denotes as the tow-photon detuning of the $i$-th channel.

Since the optical coherences $ \rho_{31}^{(i)} $ and $\rho_{32}^{(i)}$ decay much faster than the ground-state coherence $ \rho_{12}^{(i)}$, we can assume that $\rho_{33}^{(i)} = 0$, and the atoms oscillate slowly in the ground-state subspace spanned by $\{|1\rangle, |2\rangle\}$.
Furthermore, in our case that the control field is sufficiently stronger than the probe field, i.e., $\Omega_c \gg \Omega_p$, all atomic population is optically pumped into the state $|2\rangle$, leading to $\rho_{22}^{(i)} = 1$ and $\rho_{11}^{(i)} = 0$. We also have $\rho_{31}^{(i)} = 0$, since both $\rho_{11}^{(i)}$ and $ \rho_{33}^{(i)}$ vanish. 

By setting $ \dot{\rho}_{31}^{(i)} = \dot{\rho}_{32}^{(i)} = 0 $, the optical coherences can be expressed in terms of the ground-state coherence as $\rho_{31}^{(i)}=\frac{i\Omega_p^{(i)}\rho_{21}^{(i)}}{\gamma_{13} - i\Delta_c^{(i)}}\approx \frac{i\Omega_p^{(i)}\rho_{21}^{(i)}}{\gamma_{13}}$ and $\rho_{32}^{(i)}=\frac{i\Omega_c^{(i)}\rho_{12}^{(i)} + i\Omega_p^{(i)}}{\gamma_{23} - i\Delta_p^{(i)}}\approx \frac{ i\Omega_c^{(i)}\rho_{12}^{(i)}+i\Omega_p^{(i)}}{\gamma_{23}}$, where we have neglected the small single-photon detunings $\Delta_c^{(i)}$ and $\Delta_p^{(i)}$ compared to the decay rates $\gamma_{13}$ and $\gamma_{23}$. Consequently, the dynamical equation of the ground-state coherence for the $i$-th independent channel is obtained as
\begin{equation}
    \dot{\rho}_{12}^{(i)} = - \left (\gamma_{12}^{\text{eff}} + i\delta^{(i)}\right )\rho_{12}^{(i)} - \frac{\Omega_c^{(i)^{*}}\Omega_p^{(i)}}{\gamma_{23}}
\end{equation}
with the effective decay rate given by $\gamma_{12}^{\text{eff}} = \gamma_{12} + |\Omega_c^{(i)}|^2/\gamma_{23} + |\Omega_p^{(i)}|^2/\gamma_{13}$.

The ground-state coherences associated with different optical channels can be effectively coupled via the thermal motion of atoms. Following the similar procedure of Refs. \cite{2016Anti,2023Topological}, we have 
\begin{equation}
\begin{aligned}
    \dot{\rho}_{12}^{(i)} = & - \left (\gamma_{12}^{\text{eff}} + i\delta^{(i)}\right )\rho_{12}^{(i)} - \frac{\Omega_c^{(i)^{*}}\Omega_p^{(i)}}{\gamma_{23}} \\
    &+ \sum_{j\neq i}\Gamma_c^{ij} e^{-i\left [\theta_{ij}+\left(\delta^{(i)} - \delta^{(j)}\right)t\right ]}\rho_{12}^{(j)},
\end{aligned} 
\end{equation}
where the coupling rate satisfies $\Gamma_c^{ij}\propto 1/d_{ij}$, and $\theta_{ij} = \omega_{12}d_{ij}/\bar{v}$ depends on the distance between channels $ d_{ij} $ and the average velocity of the atoms $\bar{v}$. The relative phase $\theta_{ij}$ arises from the free evolution under $H_0$ as the atoms pass through the dark region between optical channels $i$ and $j$. The additional phase $\left(\delta^{(i)}-\delta^{(j)}\right)t$ originates from the transformation between channel-dependent rotating frames.

In our four-channel configuration illustrated in Fig. \ref{fig: setup}, the distances between channels are set to $d_{12} = d_{34} = d_1$ and $d_{23}=d_2$, thus the coupling rates are assumed to $\Gamma_c^{12} = \Gamma_c^{34} = v$ and $\Gamma_c^{23} = w$, where we only consider the nearest-neighbor couplings, two-photon detunings are given by $\delta^{(1)} = \Delta_A - \delta_B$, $\delta^{(2)} = \Delta_B - \delta_B$, $\delta^{(3)} = -\Delta_B - \delta_B$, $\delta^{(4)} = -\Delta_A - \delta_B$, and $\omega_{12} = -\delta_B$.
Defining the vector of the ground-state coherences as $\bm{\rho_{12}} = [\rho_{12}^{(1)}, \rho_{12}^{(2)}, \rho_{12}^{(3)}, \rho_{12}^{(4)}]^T$, we find
\begin{equation}
\bm{\dot{\rho}_{12}} = i\left [(\delta_B + i\gamma_{12}^{\text{eff}})I - H_{\text{eff}}\right ] \bm{\rho_{12}} - \bm{P},
\end{equation}
where the $i$-th component of $\bm{P}$ is $ P_i = \frac{\Omega_c^{(i)*} \Omega_p^{(i)}}{\gamma_{23}} $, and the time-dependent effective Hamiltonian is given by
\begin{equation}
H_{\text{eff}} = \begin{pmatrix}
\Delta_A & ive^{i\mu_{12}t} & 0 & 0 \\
ive^{i\mu_{21}t}& \Delta_B & iwe^{i\mu_{23}t} & 0 \\
0 & iwe^{i\mu_{32}t} & -\Delta_B & ive^{i\mu_{34}t} \\
0 & 0 & ive^{i\mu_{43}t} & -\Delta_A
\end{pmatrix}
\end{equation}
with $\mu_{ij} = -\theta_{ij} - \left(\delta^{j} - \delta^{i}\right)t$. By choosing a fixed time $t$ such that all phases $ e^{i(\delta^{(i)}-\delta^{j})t} = 1 $, the above effect Hamiltonian simplifies into the following form,
\begin{equation}
H_{\text{eff}}^{\prime} = \begin{pmatrix}
\Delta_A & ive^{i\theta_{1}} & 0 & 0 \\
ive^{i\theta_{1}}& \Delta_B & iwe^{i\theta_{2}} & 0 \\
0 & iwe^{i\theta_{2}} & -\Delta_B & ive^{i\theta_{1}} \\
0 & 0 & ive^{i\theta_{1}} & -\Delta_A
\end{pmatrix},
\end{equation}
where $\theta_i = \delta_B d_i/\bar{v}$ for $i = 1, 2$.
When the applied magnetic field is swept to zero ($\delta_B = 0$) during the EIT spectrum measurements, we arrive at the fourth-channel non-Hermitian Hamiltonian \eqref{eq: Hamiltonian}.

\section{\label{sec: noise}Noise analysis}

Although fourth-order EPs indeed exhibit superior sensing performance under ideal conditions, their practical realization requires highly precise tuning to the parameter regime at EPs. In experiments, implementation of the non-Hermitian Hamiltonian with the nearest-neighbor (NN) couplings between optical channels also suffer from the noise arising from the next-nearest-neighbor (NNN) and third-nearest-neighbor (TNN) couplings due to isotropic atomic collisions. 
To simulate the effect of such noise, we introduce a noise Hamiltonian in addition to the ideal Hamiltonian in Eq. \eqref{eq: Hamiltonian},
\begin{equation}
    H_{\text{noise}}=\left(\begin{array}{cccc}
         0&0&i\kappa&i\xi \\
         0&0&0&i\kappa\\
         i\kappa&0&0&0\\
         i\xi&i\kappa&0&0
    \end{array}\right)
\end{equation}
where $\kappa$ and $\xi$ denote the strengths of the NNN and TNN couplings, respectively, which are evaluated by Monte Carlo simulation (see Appendix \ref{sec: MCS}).

\begin{figure}
   \centering
   \includegraphics[width=1\linewidth]{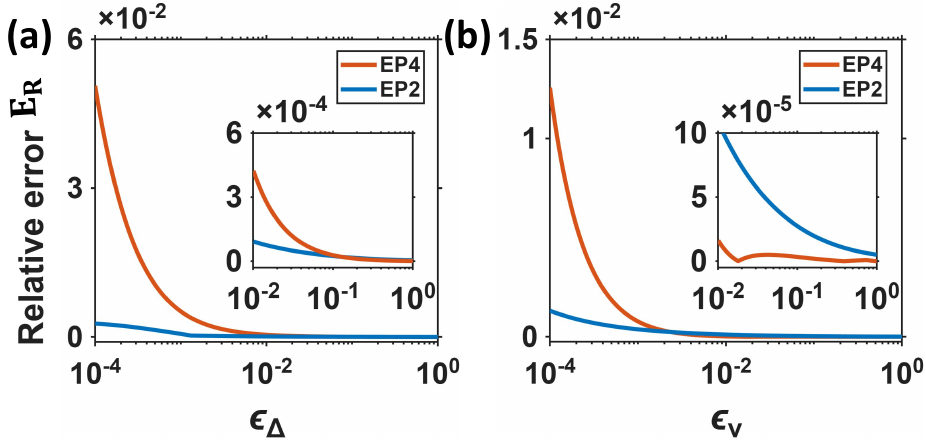}   
   \caption{Relative errors between the ideal eigenvalues and those perturbed by the NNN and TNN couplings, plotted as functions of  $\epsilon_{\Delta}$ (a) and $\epsilon_{v}$ (b). The red and blue solid curves indicate the behaviors near fourth-order and second-order EPs, respectively.}
   \label{fig: error}
\end{figure}

We calculate the relative error defined as $E_r = \left |(\lambda_{\text{noise}}-\lambda_{\text{th}})/\lambda_{\text{noise}}\right|$,
where $\lambda_{\text{noise}}$ and $\lambda_{\text{th}}$ are the maximum eigenvalues in the noisy and ideal cases. The results are plotted in Figs. \ref{fig: error}(a) and \ref{fig: error}(b) as functions of $\epsilon_{\Delta}$ and $\epsilon_v$, respectively. For small disturbances $\epsilon_{\Delta}$ and $\epsilon_{v}$, the relative error at a second-order EP is significantly smaller than that at a fourth-order EP. This indicates that although the fourth-order EP offers higher sensitivity enhancement, it is also more susceptible to noise; in contrast, the second-order EP demonstrates stronger robustness against noise, despite its lower sensitivity. 

\section{\label{sec: MCS}Evaluation of Next-Nearest-Neighbor and Third-Nearest-Neighbor Coupling strengths}

\begin{figure}
    \centering
    \includegraphics[width=1\linewidth]{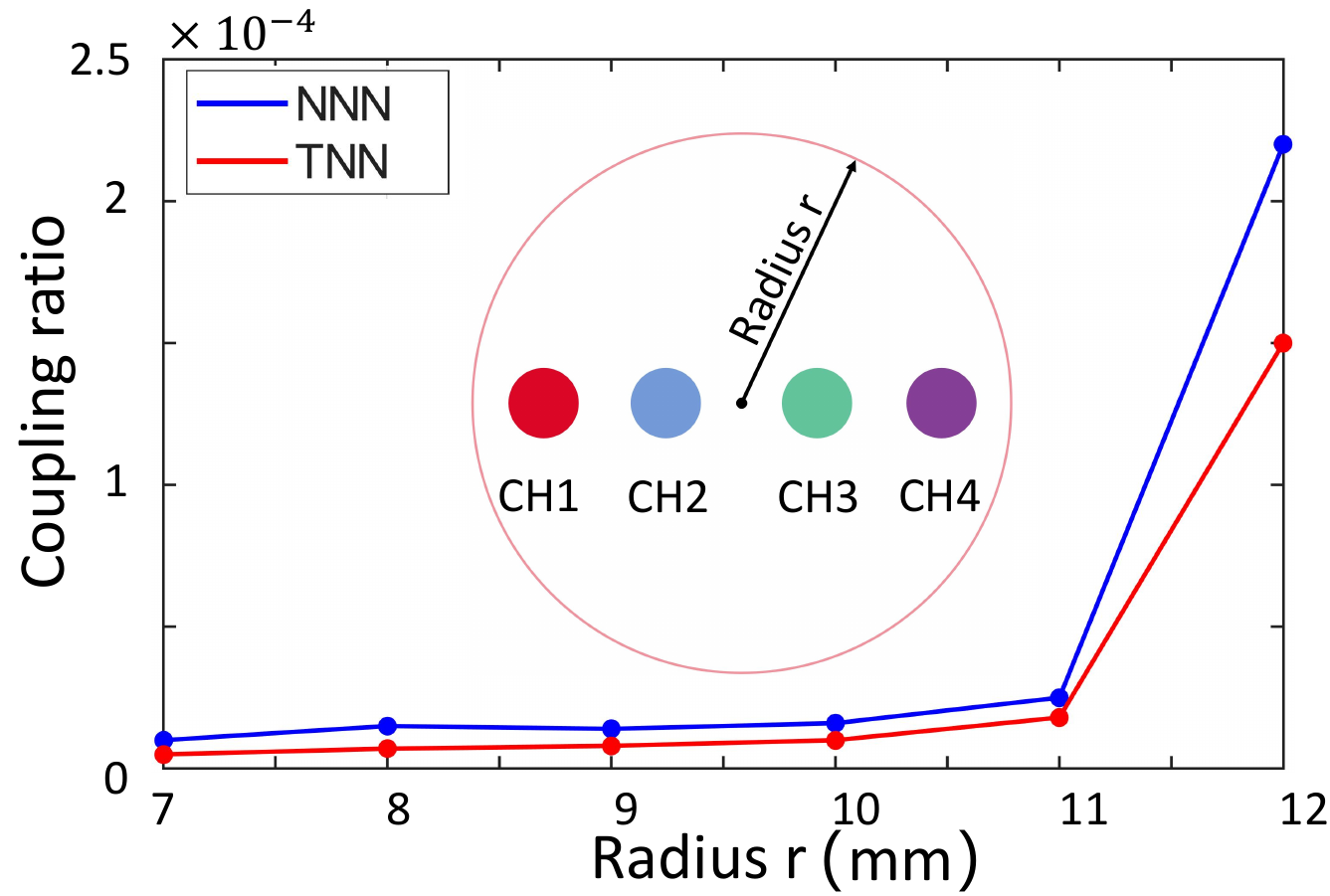}
    \caption{Ratios of the NNN (blue) and TNN (red) couplings to the NN couplings is plotted as a function of the ring-shaped pump beam radius $r$. The simulation tracked \( 3 \times 10^6 \) atoms, each for \( 10^5 \) steps. Inset is the schematic of a ring-shaped laser beam (pink) close to the cell wall with radius $r$.}
    \label{fig: nnncoupling}
\end{figure}

Owing to the thermal motion of atoms, the ground-state coherence can be effectively coupled not only via the nearest-neighbor (NN) optical channels, but also through the next-nearest-neighbor (NNN) and the third-nearest-neighbor (TNN) channels.
In practical implementations, these undesired couplings beyond the NN channels can be largely suppressed by using a ring-shaped laser beam with radius $r$ placed close to the cell wall (see the inset of Fig. \ref{fig: nnncoupling}) and optically pumping the atoms back to a thermal steady state. 
For simulating the effects of the ring-shaped pumping beam, we employ the Monte Carlo method~\cite{2023Topological} to estimate the strengths of the NNN and TNN couplings. 

The mean free path of the atomic collisions is characterized by
\begin{equation}
l_0 = \frac{k_BT}{\sqrt{2}\pi r_a^2 P}\approx 30 m
\end{equation}
where $k_B$ is the Boltzmann constant, $T$ is the temperature, $r_a$ is the effective atomic radius, $P$ is the cell pressure.
Since the mean free path is significantly larger than the cell radius ($12.5$ mm), the probability of atomic collisions is extremely small. Consequently, unlike the direct NN couplings, the NNN and TNN couplings are mainly induced by the isotropic motions of atomic collisions.

Figure \ref{fig: nnncoupling} plots the the results of Monte Carlo simulations, showing the ratios of the NNN and TNN couplings to the NN coupling as functions of the radius $r$ of the ring-shaped optical pumping beam. 
When $r = 12$ mm, the NNN coupling is stronger than the TNN coupling because the NNN separation is shorter. Importantly, both noise contributions can be effectively suppressed by decreasing $r$, thereby reducing the likelihood of atoms transporting coherence to farther-than-nearest neighboring channels.

\nocite{*}

\bibliography{Refs_NHS}

\end{document}